\documentclass[reprint,showpacs,amssymb,amsmath,aps]{revtex4-1}
\usepackage{graphics}
\usepackage{bm}
\usepackage{amsthm}
\begin{document}

\title{Non-transverse factorizing fields and entanglement in finite spin systems}
\author{M. Cerezo, R. Rossignoli, N. Canosa}
\affiliation{Departamento de F\'{\i}sica-IFLP,
Universidad Nacional de La Plata, C.C.67, La Plata (1900), Argentina}
\begin{abstract}

We determine the conditions for the existence of non-transverse factorizing
magnetic fields in general spin arrays with anisotropic $XYZ$ couplings of
arbitrary range. It is first shown that a uniform maximally aligned completely
separable eigenstate can exist just for fields $\bm{h}_s$ parallel to a
principal plane and forming four straight lines in field space, with the
alignment direction different from that of $\bm{h}_s$ and determined by the
anisotropy. Such state always becomes a non-degenerate ground state (GS) for
sufficiently strong (yet finite) fields along these lines, in both
ferromagnetic (FM) and antiferromagnetic (AFM) type systems. In AFM chains,
this field coexists with the non-transverse factorizing field $\bm{h}'_s$
associated with a degenerate N\'eel-type separable GS, which is
shown to arise at a level crossing in a finite chain. It is also demonstrated
for arbitrary spin that pairwise entanglement reaches full range in the
vicinity of both $\bm{h}_s$ and $\bm{h}'_s$, vanishing  at $\bm{h}_s$ but
approaching small yet finite side-limits at $\bm{h}'_s$, which are analytically
determined. The behavior of the block entropy and entanglement spectrum in
their vicinity is also analyzed.
\end{abstract}
\maketitle

\section{Introduction}
The ground state (GS) of strongly interacting spin systems immersed in a
magnetic field $\bm{h}$ can exhibit, under certain conditions, the remarkable
phenomenon of factorization \cite{Ku.82}, i.e., of becoming a product of single
spin states. Such exact factorization can occur at finite fields despite the
strong couplings existing between the spins, albeit at very specific values
(and orientation) of the field. In the seminal work of ref.\ \cite{Ku.82}, it
was shown that antiferromagnetic (AFM) chains with first neighbor $XYZ$
couplings possess a separable N\'eel-type GS (NGS) if the field vector lies on
the surface of an ellipsoid determined by the couplings. Factorization was then
investigated in other models with transverse fields
\cite{Mu.85,TR.04,TR.05,Am.06,FB.07,GI.07,GI.08,GI.09,RCM.08,RCM.09,G.09,
GI.10,CRM.10,CRC.10,ARL.10,TR.11,SC.13,GM.13,KC.14}, with a general
formalism for describing factorization introduced and discussed in
\cite{GI.07,GI.08,GI.09}.

In \cite{RCM.08,RCM.09,CRM.10} we have shown that in {\it finite} $XYZ$ chains,
the transverse factorizing field (TFF) $\bm{h}_{zs}$ pointing along a principal
axis ($z$) corresponds actually to the last GS $S_z$ parity transition (level
crossing). The ensuing separable GS is two-fold degenerate, breaking a basic
symmetry of the Hamiltonian ($S_z$ parity). The non-transverse factorizing
fields (NTFF) $\bm{h}'_s$ of \cite{Ku.82} will be shown to also belong to this
class in finite cyclic chains, i.e., they arise at the last GS level crossing
and determine a degenerate separable GS breaking translational invariance (TI).
In finite systems the underlying mechanism of factorization in these cases is
the existence of separable linear combinations of the symmetry preserving
entangled crossing states.

In this work we first determine the general conditions for exact factorization
under non-transverse fields. It is then shown that a {\it uniform}
non-degenerate separable GS (UGS) does exist at a field $\bm{h}_s$ which does
not belong in general to the ellipsoid of ref.\ \cite{Ku.82}, and does not
correspond to a level crossing. This GS actually arises in both AFM and FM-type
systems, even for couplings of arbitrary range provided there is a fixed
anisotropy ratio, but only for fields parallel to a principal plane, with the
set of fields $\bm{h}_s$ forming four straight lines. Factorization emerges
here from the splitting of the degenerate separable eigenstates existing at the
TFF $\bm{h}_{zs}$. Unlike $\bm{h}'_s$, $\bm{h}_s$ can be arbitrarily strong,
allowing  the separation of the UGS from the remaining spectrum. This enables
an easy preparation of an exactly separable state, which can be useful for
quantum information applications (a product initial state is assumed in the
standard model of quantum computation \cite{NC.00}).

A second but not less important aspect of factorization is that it corresponds
to an {\it entanglement transition}: In the transverse case, the factorizing
field is, remarkably,  the point where pairwise entanglement reaches {\it full
range} in its immediate vicinity, and changes its type
\cite{Am.06,FB.07,RCM.08,RCM.09,CRM.10}. We had previously shown that the
entanglement between any two spins reaches in a finite chain weak yet finite
{\it common side-limits} at the transverse field $\bm{h}_{zs}$, irrespective of
separation or coupling range \cite{RCM.08,RCM.09}, arising from the entangled
crossing states. This type of limit  also occurs at the NTFF $\bm{h}'_s$ of
ref.\  \cite{Ku.82}, as will be shown. But in addition, we will prove that
pairwise entanglement also reaches full range at the vicinity of the NTFF
$\bm{h}_s$ leading to a non-degenerate UGS. Here the entanglement between any
two spins, though 0 at $\bm{h}_s$,  is turned on as $\bm{h}_s$ is approached
from either side, with the concurrence vanishing  then linearly with
$|\bm{h}-\bm{h}_s|$. The underlying reason is essentially the monogamy of
entanglement \cite{CKW.00,OV.06}, which prevents distant pairs from becoming
entangled if first or close neighbors are strongly entangled.  In the vicinity
of  $\bm{h}_{s}$, close neighbor entanglement decreases strongly, allowing the
emergence of weak yet non-zero entanglement between distant pairs. The behavior
of the block entanglement entropy in the vicinity of the  NTFF will be also
analyzed. It will be shown to vanish essentially quadratically at $\bm{h}_s$,
while  at $\bm{h}'_s$ it will approach finite side-limits in a finite chain,
which will be analytically determined. The entanglement spectrum will indicate,
as expected, just one nonzero eigenvalue at $\bm{h}_s$, although at the
side-limits of $\bm{h}'_s$ two nonzero eigenvalues will remain.

The general equations for non-transverse factorizing fields and its uniform 
and N\'eel-type solutions are derived and discussed in section \ref{II}, 
whereas entanglement together with illustrative results for the pairwise 
concurrence, block entropy, entanglement spectrum and magnetization in FM 
and AFM chains with $XY$ and $XYZ$ couplings under non-transverse fields 
are discussed in section III. Conclusions are derived in IV.

\section{Factorization in non-transverse fields\label{II}}
\subsection{General Equations}
We consider an array of $n$ spins $S_i$ not necessarily equal, interacting
through $XYZ$ Heisenberg couplings of arbitrary range in the presence of a
general magnetic field $\bm{h}^i=(h^i_x,h^i_y,h^i_z)$, not necessarily uniform.
The Hamiltonian reads
\begin{equation}
H=-\sum_{i,\mu} h^i_\mu S^\mu_i - {\textstyle\frac{1}{2}} \sum_{i\neq j,\mu} J_\mu^{ij}
S^\mu_i S^\mu_j\,, \label{H}
\end{equation}
where $i,j$ label the sites in the array, $S_i^\mu$, $\mu=x,y,z$, the spin
components at site $i$ and $J_\mu^{ij}$  the coupling strengths between spins
$i$ and $j$ ($J_\mu^{ij}\geq0$ corresponds to the FM case whilst
$J_\mu^{ij}\leq0$ to the AFM case). In the transverse case $h^i_x=h^i_y=0$
$\forall$ $i$, $H$ conserves the $S_z$-parity
$P_z=\exp[\imath\pi\sum_i(S^z_i+S_i)]$ ($[H,P_z]=0$). This symmetry no longer
holds for non-transverse fields.

We now determine the general conditions for which $H$ possesses a  {\it
completely separable eigenstate} of the form
\begin{equation}|\Theta\rangle=\otimes_{i=1}^n
R_i|0_i\rangle,\;\;R_i=\exp[-\imath \phi_i S^z_i]\exp[-\imath \theta_i
S^y_i]\,, \label{Theta}
 \end{equation}
where $|0_i\rangle$ denotes the local state with maximum spin along $z$
($S^z_i|0_i\rangle=S_i|0_i\rangle$) and $R_i$ rotates this state to direction
$\bm{n}_{i}=(\sin\theta_i\cos\phi_i,\sin\theta_i\sin\phi_i,\cos\theta_i)$. The
equation $H|\Theta\rangle=E_\Theta|\Theta\rangle$  leads, after writing $H$ in
terms of  the rotated spins $S_i^{\mu'}=R_i S_i^\mu R_i^\dagger$, to:
\\
I) The field independent equations
\begin{eqnarray}
&&J_y^{ij}(\cos\phi_i\cos\phi_j-\cos\theta_i \sin\phi_i \cos\theta_j\sin\phi_j)=\nonumber\\
&&J_x^{ij}(\cos\theta_i \cos\phi_i \cos\theta_j \cos\phi_j-\sin\phi_i \sin\phi_j)\nonumber\\&&
+J_z^{ij} \sin\theta_i \sin\theta_j\,, \label{1}\\
&&J_y^{ij}(\cos\theta_i \sin\phi_i \cos\phi_j+\cos\phi_i \cos\theta_j \sin\phi_j)=\nonumber\\
&&J_x^{ij}(\cos\theta_i \cos\phi_i \sin\phi_j+\sin\phi_i \cos\theta_j \cos\phi_j)\,,\label{2}
\end{eqnarray}
which are also independent of spin and are responsible for
cancelling all elements connecting $|\Theta\rangle$ with two-spin excitations,  and \\
II) The field dependent equations
\begin{eqnarray}
&& h^i_z \sin\theta_i-\cos\theta_i(h^i_x\cos\phi_i+h^i_y\sin\phi_i)=
\nonumber\\&&{\textstyle\sum\limits_{j\neq i}}
S_j[\cos\theta_i \sin\theta_j (J_x^{ij}\cos\phi_i\cos\phi_j+J_y^{ij}\sin\phi_i \sin\phi_j)\nonumber\\
&&-J_z^{ij}\sin\theta_i \cos\theta_j]\label{3}\,,\\
&&h^i_x \sin\phi_i -h^i_y\cos\phi_i=\nonumber\\
&&{\textstyle\sum\limits_{j\neq i}}S_j\sin\theta_j[-J_x^{ij}\sin\phi_i
 \cos\phi_j+J^{ij}_y \cos\phi_i \sin\phi_j]\,,\label{4}
\end{eqnarray}
which cancel all elements connecting $|\Theta\rangle$ with single spin
excitations and are just the  mean field  stationary equations
$\partial_{\theta_i}\langle H\rangle=0$, $\partial_{\phi_i}\langle H\rangle=0$,
where
\begin{equation}
\langle H\rangle\equiv\langle \Theta|H|\Theta\rangle= -\sum_{i}\langle
\bm{S}_i\rangle\cdot(\bm{h^i}+{\textstyle\frac{1}{2}}\sum_j
 \bm{J}^{ij}\langle\bm{S}_j\rangle)\,,\label{EE}\end{equation}
with $\langle \bm{S}_i\rangle=S_i\bm{n}_{i}$ and $\bm{J}^{ij}$ a diagonal
matrix of elements $J^{ij}_{\mu}$. If Eqs. (\ref{1})--(\ref{2}) are satisfied $\forall$
$i,j$, Eqs.\ (\ref{3})--(\ref{4}) determine the set of {\it factorizing
fields}.

In terms of the alignment directions $\bm{n}_i\equiv \bm{n}_i^{z'}$ and 
the orthogonal unit vectors $\bm{n}_i^{y'}=(-\sin\phi_i,\cos\phi_i,0)$,
$\bm{n}_i^{x'}=(\cos\theta_i\cos\phi_i,\cos\theta_i\sin\phi_i,-\sin\theta_i)$, 
 we may also express Eqs.\
(\ref{1})--(\ref{2}) as
\begin{eqnarray}
\bm{n}^{x'}_i\cdot\bm{J}^{ij}\bm{n}^{x'}_j&=&\bm{n}^{y'}_i\cdot\bm{J}^{ij}
\bm{n}^{y'}_j\,,\label{11}\\
\bm{n}^{x'}_i\cdot\bm{J}^{ij}\bm{n}^{y'}_j&=&-\bm{n}^{y'}_i\cdot\bm{J}^{ij}\bm{n}^{x'}_j
\,,\label{22}
\end{eqnarray}
which imply $J^{ij}_{x'x'}=J^{ij}_{y'y'}$ and $J^{ij}_{x'y'}=-J^{ij}_{y'x'}$
when writing the coupling in (\ref{H}) in terms of the rotated spins $S_i^{\mu'}$,  i.e.,
$\sum_\mu J^{ij}_{\mu}S_i^\mu S_j^\mu=\sum_{\mu,\nu}
{J}^{ij}_{\mu'\nu'}S_i^{\mu'} S_j^{\nu'}$. And Eqs.\ (\ref{3})--(\ref{4}) become
\begin{eqnarray}
\bm{n}_i^{\mu'}\cdot(\bm{h}^i+ \sum_j
\bm{J}^{ij}\langle\bm{S}_j\rangle)=0,\;\;\mu'=x',y'\,, \label{33}
\end{eqnarray}
implying that $\bm{h}^i$ should cancel the components of $\sum_j
\bm{J}^{ij}\langle\bm{S}_j\rangle$ orthogonal to the alignment direction, such
that \[\bm{h}^i+\sum_j\bm{J}^{ij}\langle\bm{S}_j\rangle\propto\bm{n}_i\,.\] The
general solution for the NTFF at site $i$ is then
\begin{equation}
\bm{h}^i_s=\bm{h}^i_{\parallel}+\bm{h}^i_{\perp}\,,    \label{hpp}
\end{equation}
where $\bm{h}^i_{\parallel}=h^i_{\parallel}\bm{n}_i$ is  an {\it arbitrary}
field {\it  parallel} to the local alignment direction, which just shifts the
energy (\ref{EE}),  and
\begin{equation}
\bm{h}^i_{\perp}=-\sum_{j}[\bm{J}^{ij}\langle\bm{S}_j\rangle
-\bm{n}_i(\bm{n}_i\cdot\bm{J}^{ij}\langle\bm{S}_j\rangle)]\,,\label{hp}
\end{equation}
is a field {\it orthogonal} to the alignment direction ($\bm{n}_i\cdot\bm{h}^i_{\perp}=0$),
representing the NTFF
of {\it lowest} magnitude. Nonetheless, a finite $h^i_{\parallel}$ will be
normally  required in order that $|\Theta\rangle$ be a GS (see sec.\ II.C). Let
us remark, finally, that Eqs.\ (\ref{11})--(\ref{hp}) remain valid for general
couplings $\sum_{\mu,\nu}J^{ij}_{\mu\nu}S_i^\mu S_j^\nu$ in (\ref{H}).

\subsection{Uniform solution \label{IIb}}
\begin{figure}[t]
\centerline{\scalebox{.7}{\includegraphics{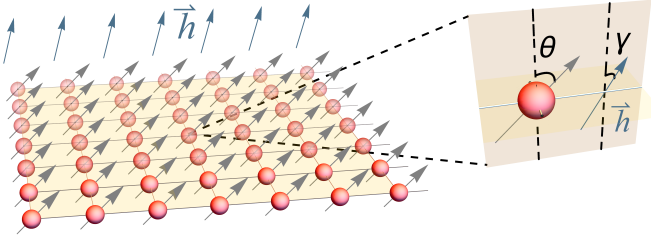}}}
\caption{(Color online) Schematic plot of the uniform solution.}
 \label{f1}
\end{figure}

Eqs.\ (\ref{1})--(\ref{4}) (or (\ref{11})--(\ref{33})) are quite
general and describe a wide range of interesting scenarios. We examine first
the possibility of a {\it uniform} solution with $\theta_i=\theta$,
$\phi_i=\phi$ $\forall$ $i$ (Fig.\ \ref{f1}), such that $|\Theta\rangle$ is a
maximum spin state: $|\langle\sum_i \bm{S}_i\rangle|=\sum_i S_i$. Such
solution preserves TI and then has the possibility to be a non-degenerate GS in
systems with this invariance under a uniform field.

Eq.\ (\ref{2}) becomes $(J_x^{ij}-J_y^{ij})\cos\theta\sin2\phi=0$, implying, if
$J_x^{ij}-J_y^{ij}\neq 0$ for at least one pair, that the spin vector
$\langle\bm{S}_i\rangle$ should be {\it parallel to a principal plane} ($xz$ if
$\phi=0$, $yz$ if  $\phi=\pi/2$ and $xy$ if $\theta=\pi/2$). Without loss of
generality, we can assume $\phi=0$ (the other choices are rotations of this
case). Eq.\ (\ref{1}) then leads to
\begin{equation}
\cos^2\theta=\frac{J_y^{ij}-J_z^{ij}}{J_x^{ij}-J_z^{ij}}=\chi\,,\label{7}
\end{equation}
if $J_x^{ij}\neq J_z^{ij}$, implying a {\it constant anisotropy ratio} $\chi$ for
these pairs, and an isotropic coupling $J^{ij}_\mu=J^{ij}$ $\forall$ $\mu$ if
$J_x^{ij}=J_z^{ij}$. The condition $0\leq \chi\leq 1$ imposes the restriction
\begin{equation}
J_x^{ij}\geq J_y^{ij}\geq J_z^{ij}\;\;{\rm or}\;\;J_x^{ij}\leq J_y^{ij}\leq
J_z^{ij}\label{8}\,.
\end{equation}
Eqs.\ (\ref{7})--(\ref{8}) entail that  the
$J_\mu^{ij}$ should be of the form
\begin{equation}
J_\mu^{ij}=J^{ij}+r^{ij}J_\mu\,,\label{rij}
\end{equation}
with the  $J_\mu$'s satisfying (\ref{8}). The state $|\Theta\rangle$ will then
depend just on $\chi=\frac{J_y-J_z}{J_x-J_z}$, being {\it independent of the
coupling range} determined by $J^{ij}$ and $r^{ij}$. Notice that Eq.\ (\ref{7})
leads to four possible alignment directions in the $xz$ plane, corresponding to
the solutions $\pm\theta\,$ and $\pm(\pi-\theta)$, with $\theta\in(0,\pi/2)$.

We remark that in the fully isotropic case $r_{ij}=0$ $\forall$ $i,j$ in
(\ref{rij}) (rotationally invariant coupling), $\theta$ and $\phi$ remain
obviously arbitrary under Eqs.\ (\ref{1})--(\ref{2}), whereas in the $XX$ case
$J^{ij}_x=J_y^{ij}$ $\forall$ $ij$ (coupling invariant under any rotation around
the $z$ axis), Eq.\ (\ref{2}) is trivially satisfied while (\ref{1}) leads to
$\sin\theta=0$ if $J_x^{ij}\neq J_z^{ij}$ for at least one pair, in agreement
with (\ref{7}), implying alignment just in the $z$ direction. We will focus in
what follows on the anisotropic case $0<\chi<1$, where the alignment direction
is non-trivial ($\theta\in(0,\pi/2)$).

\begin{figure}[t]
 \centerline{\scalebox{.7}{\includegraphics{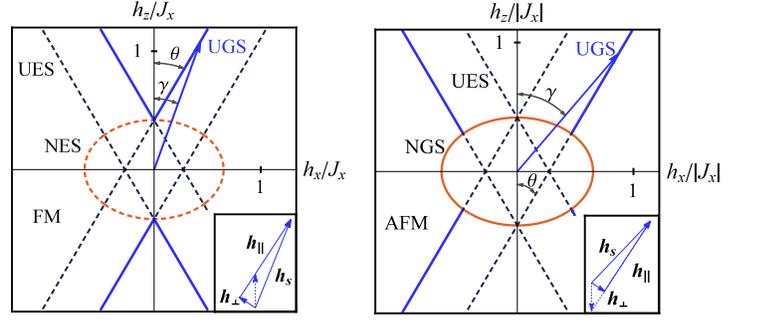}}}
\caption{(Color online) Factorizing fields for ferromagnetic (left) and
antiferromagnetic (right) $XYZ$ chains in the $xz$ principal plane of the field
space. Solid straight  lines depict the fields determining a uniform ground
state (UGS), whereas the dashed straight lines those determining uniform
excited eigenstates (UES). The ellipse depicts the fields corresponding to the
N\'eel-type ground state (NGS, solid lines) or excited eigenstate (NES, dashed
line). The plot corresponds to $J_z=0$ and $J_x>0$ ($<0$) in the FM (AFM) case,
with $0<J_y/J_x<1$. The arrow indicates a direction of the external field along
which one (FM) or two (AFM) GS factorizing fields are encountered as its
magnitude increases. The  field direction $\bm{n}_\gamma$ differs from the spin
alignment direction $\bm{n}_\theta$. The insets depict the decomposition
(\ref{91}) of the non-transverse factorizing field  for the UGS in both
diagrams, with the dashed arrow indicating the transverse factorizing field
$\bm{h}_{zs}$.} \label{f2}
\end{figure}

For $\phi=0$, Eqs.\ (\ref{4}) (or (\ref{33})) imply
$h_y^i=0$, i.e., the field at each site {\it should be  parallel to the
corresponding principal plane} ($xz$). Eq.\ (\ref{3}) then becomes
\begin{eqnarray}
h_z^i\sin\theta -h_x^i\cos\theta =h^i_{\perp},\label{EcBu}\end{eqnarray}
 where
\begin{eqnarray}
 h^i_{\perp}&=&
\sin\theta\cos\theta\sum_{j\neq i}S_j(J_x^{ij}-J_z^{ij})\,. \label{EcBu2}
\end{eqnarray}
Setting $\bm{n}_\theta=\bm{n}_i=(\sin\theta,0,\cos\theta)$  and
$\bm{n}^\perp_\theta=(-\cos\theta,0,\sin\theta)$, Eqs.\
(\ref{EcBu})--(\ref{EcBu2}) imply that the NTFF is given by
 \begin{eqnarray}
 \bm{h}_s^i&=&h_{\parallel}^i\bm{n}_\theta+h_{\perp}^i\bm{n}_\theta^\perp\,,\label{91}
\end{eqnarray}
in agreement with (\ref{hpp}), with $h^i_{\parallel}$ arbitrary and
$h_{\perp}^i\bm{n}_\theta^\perp$ orthogonal to the alignment direction.   Eqs.\
(\ref{EcBu2})--(\ref{91}) give rise to a family of NTFF {\it lying along four
straight lines} (Fig.\ \ref{f2}), one for each alignment direction.

Note that field and spin directions {\it cannot be parallel} if
$h^i_{\perp}\neq 0$: At fixed field direction
$\bm{n}_\gamma=(\sin\gamma,0,\cos\gamma)$, i.e.,
$\bm{h}^i_s=h^i_s(\gamma)\bm{n}_\gamma$,  Eqs.\ (\ref{91})  leads to
\begin{equation}
h^i_{s}(\gamma)=\frac{h^i_\perp}{\sin(\theta-\gamma)}
\label{10}\,,
\end{equation}
which diverges for $\gamma\rightarrow\theta$.  When the four values of $\theta$
are considered, Eq.\ (\ref{10}) leads to two distinct values of $|\bm{h}^i_s|$
at fixed $\gamma\neq\pm\theta$, which merge at the principal axes (Fig.\
\ref{f2}).

For $\gamma=0$, we recover from (\ref{10}) the TFF \cite{RCM.08,RCM.09}
\begin{equation}
h^i_{zs}=h^i_s(0)=\frac{h^i_{\perp}}{\sin\theta}\,,
\end{equation}
which is  the solution of (\ref{EcBu}) for $h^i_x=0$.
We can then also express Eq.\ (\ref{91}) as ($\bm{n}_z=(0,0,1)$)
\begin{eqnarray}
\bm{h}^i_s&=&h^i\bm{n}_\theta+h^i_{zs}\bm{n}_z\,,\label{9}
\end{eqnarray}
where $h^i=h^i_{\parallel}-h^i_{\perp}/\tan\theta$. Hence, we can also consider
$\bm{h}^i_s$ as the sum of the TFF $\bm{h}^i_{zs}=h^i_{zs}\bm{n}_z$ plus a
non-transverse field {of arbitrary magnitude} $h_i$ along the spin alignment
direction $\bm{n}_\theta$, which just shifts the energy $E_\Theta$.

In systems with TI (i.e., infinite or cyclic), $S_i=S$ and
$h^i_{\perp}=h_{\perp}$ $\forall$ $i$, implying a {\it uniform} factorizing
field $h_s(\gamma)$ at fixed orientation $\gamma$. Nonetheless, Eqs.\
(\ref{9})--(\ref{10}) show that the uniform solution remains feasible even in
the absence of TI, provided the $h^i_\mu$ at each site can be controlled
independently. In particular, in open finite uniform chains or lattices with
short range couplings, the uniform separable solution requires just border
corrections to the otherwise uniform bulk factorizing field.

\subsection{Uniform ground state}

For the uniform solution, the energy (\ref{EE}) becomes
\begin{eqnarray}
E_\Theta
 &=&-{\textstyle\frac{1}{2}}\sum_{i,j}S_i
 S_j(J_x^{ij}-J_y^{ij}+J_z^{ij})-\sum_i S_i h^i_{\parallel}\,\label{E2}\\
&=&-{\textstyle\frac{1}{2}}\sum_{i,j}S_i
 S_j(J_x^{ij}+J_y^{ij}-J_z^{ij})-\sum_i S_i h^i\,.\label{E}
\end{eqnarray}
It is then apparent that $|\Theta\rangle$ {\it will  be  GS  if the fields
$h^i_{\parallel}\bm{n}_{\theta}$ (or equivalently $h^i\bm{n}_{\theta}$) along
the spin alignment direction are sufficiently strong},  since no other  state
has an energy which decreases more rapidly with the applied field. Therefore, a
{\it transition} to this uniform separable GS (UGS) will always arise as
$h^i_{\parallel}$ increases, in {\it both} FM or AFM-type systems, as can be
appreciated in Fig.\ \ref{f2} (transition from dashed to solid along the
straight lines). Before this transition, $|\Theta\rangle$ is an excited
eigenstate (no other state can increase its energy more rapidly for decreasing
$h^i$'s).

We now show that if Eq.\ (\ref{7}) is satisfied and, $\forall$ $i,j$,
\begin{equation}
J_x^{ij}\geq |J_y^{ij}|\,,\label{14}
\end{equation}
such transition occurs {\it at the TFF} $\bm{h}^{i}_{zs}$, i.e.,
$|\Theta\rangle$ will be GS
$\forall$  $h_i\geq 0$  in (\ref{9}) (left panel in Fig.\ \ref{f2}). \\
{\it Proof:}  We first note that if $\phi_i=0$ and $\theta_i=\theta$ $\forall$ $i$,
Eq.\ (\ref{Theta}) leads to \\$|\Theta\rangle=\otimes_{i}(\sum_{k=0}^{2S_i}
\binom{2S_i}{k}^{1/2}\cos^{2S_i-k}\frac{\theta}{2}\sin^k\frac{\theta}{2}|k_i\rangle)$,
where  $S_i^z|k_i\rangle=(S_i-k)|k_i\rangle$. Eq.\ (\ref{14}) implies that the
interaction in $H$ will contain just negative or zero off-diagonal elements in
the standard basis $\{\otimes_i|k_i\rangle\}$, as seen by writing (\ref{H}) in
terms of $S_i^{\pm}=S_i^x\pm i S_i^y$. The same holds for $H$ if $h_y^i=0$ and
$h_x^i\geq 0$ $\forall$ $i$. A GS with expansion coefficients real and of
the same sign in this basis will then exist, as different signs will not
decrease $\langle H\rangle$. But such GS cannot be orthogonal to
$|\Theta\rangle$  if  $\theta\in(0,\pi)$ (implying
$h^i\geq 0$ in (\ref{9}) if $h^i_x\geq 0$), so that it must coincide with
$|\Theta\rangle$ when $|\Theta\rangle$ is an exact eigenstate. The case
$h^i_x\leq 0$ can be reduced to the previous one by a rotation of angle $\pi$
around the $z$ axis, which leaves the rest of $H$ unchanged.

Besides, in the transverse case $h_i=0$ $\forall i$, the states
$|\Theta\rangle$ and $|-\Theta\rangle=P_z|\Theta\rangle$, obtained for
$\theta=\pm|\theta|$, become degenerate (Eq.\ (\ref{E})). The TFF
$\bm{h}^i_{zs}$ determines then a pair of degenerate UGS $|\pm\Theta\rangle$
when (\ref{14}) holds \cite{RCM.08}, and the addition of a field parallel to
$\bm{n}_{\theta}$ ($\bm{n}_{-\theta}$) removes this degeneracy, leaving just
$|\Theta\rangle$ ($|-\Theta\rangle$) as GS. The transition to the UGS takes
then place at $\bm{h}^i_{zs}$. \qed

The gap to the first excited state can then be made arbitrarily large by
increasing the fields $h_i$ (Eq.\ (\ref{E})). Note that the similar case
$J_z^{ij}\geq |J_y^{ij}|$ $\forall$ $i,j$ can be reduced to the previous one
after a $\pi/2$ rotation around the $y$ axis. Hence, in this case the
transition takes place at the transverse field along $x$,
$h^i_{xs}=h^i_{\perp}/\cos\theta=h^i_{zs}\tan\theta$.

\subsection{N\'eel-type solutions.}
In addition to the  uniform solution,  other solutions of Eqs.\
(\ref{1})--(\ref{4}) can exist, which break TI. This is the case of the
N\'eel-type separable eigenstates determined in \cite{Ku.82} for the AFM chain
with first neighbor couplings in a uniform field ($J_{ij}=0$,
$r^{ij}=\delta_{i,j\pm 1}$ in (\ref{8}), with $J_\mu\leq 0$ for $\mu=x,y,z$),
where $\theta_i,\phi_i$ have {\it alternating} values. In a finite cyclic chain
(with an {\it even} number $n$ of spins), such solution  must then be two-fold
degenerate, arising  at the {\it crossing of two non-separable TI eigenstates}.
The mechanism is then similar to that of the TFF for the uniform solution
\cite{RCM.08}. The associated NTFF $\bm{h}'_s$ points to the surface of an
ellipsoid \cite{Ku.82}, given for $S=1/2$ by
 \begin{eqnarray}
 \frac{{h'_s}_{x}^2}{(J_x+J_y)(J_x+J_z)}+
 \frac{{h'_s}_{y}^2}{(J_y+J_z)(J_y+J_x)}+&&\nonumber\\\frac{{h'_s}_{z}^2}{(J_z+J_x)(J_z+J_y)}&=&1\,.
 \label{Kur}
 \end{eqnarray}
Within the $xz$ plane,
$\bm{h}'_s=h'_s(\gamma)\bm{n}_\gamma$ describes an  ellipse (Fig.\ \ref{f2}),
satisfying \begin{equation}
|h'_s(\gamma)|^2=\frac{(J_x+J_z)(J_x+J_y)(J_z+J_y)}
{(J_x+J_y)\cos^2\gamma+(J_z+J_y)\sin^2\gamma}\,.\label{hgp}
\end{equation}
While in a FM-type chain such solution also exists but corresponds to an excited
eigenstate (left panel in Fig.\ \ref{f2}), in the AFM case it is a GS which
coexists with the previous UGS in the $xz$ field plane (right panel). For instance, 
they can arise for the same field orientation at different field magnitudes. This possibility 
is related with the existence of different solutions for the local unitary operations 
which can leave an eigenstate invariant in the treatment of \cite{GI.07,GI.08,GI.09}. Moreover,
the point where the straight line of the uniform solution crosses the ellipsoid
($\bm{h}_s=\bm{h}'_s$) is precisely that beyond which the uniform solution
becomes GS (at this point the N\'eel-type solution becomes uniform, coinciding
with the UGS). Hence, within the first quadrant, the UGS arises for field
angles $0\leq \gamma<\theta$ in the FM case but $\theta<\gamma\leq\gamma_{\rm
m}$ in the AFM case, with
\begin{equation}
\tan\gamma_m=\frac{J_x+J_y}{J_y+J_z}\tan\theta\,. \label{tang}
\end{equation}
Within this window, the GS of the AFM chain exhibits then {\it two distinct
factorizing fields} as $\bm{h}$ increases at fixed $\bm{n}_\gamma$ (right panel
in Fig.\ \ref{f2}), a result which has not yet been reported.

\begin{figure}[t]
 \centerline{\scalebox{.9}{\includegraphics{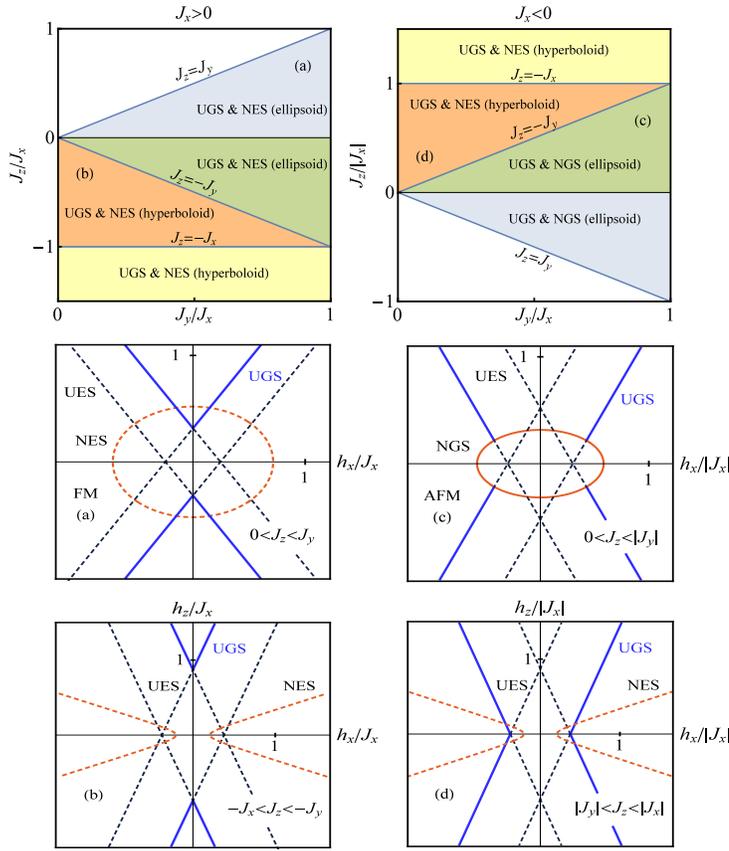}}}
\caption{(Color online) Factorization diagram (top panels) and factorizing
fields (central and bottom panels) for $XYZ$ couplings satisfying Eq.\
(\ref{8}), with $J_x>0$ ($<0$) in the left (right) panels and $J_y$ of the same
sign as $J_x$.  The different combinations of couplings are indicated (see also
text). For $J_x<J_y<0$, the N\'eel separable eigenstate ceases to be GS when
$J_z\geq -J_y$, although the uniform separable eigenstate remains GS for
appropriate fields, as seen in the top and bottom right panels. Eq.\
(\ref{Kur}) may  determine an hyperboloid when the couplings have different
signs, as seen in the top and bottom panels.}
 \label{f3}
\end{figure}

Eq.\ (\ref{Kur}) may also determine an {\it hyperboloid} when  $J_x$, $J_y$ and
$J_z$ do not have all the same signs, as shown  in Fig.\ \ref{f3}, where all
(non-equivalent) possible combinations of couplings for the case (\ref{8}) are
considered. When $|J_z|$ increases from $0$, the diagrams of Fig.\ \ref{f2}
remain essentially unchanged if $|J_z|< |J_y|$ (central panels in Fig.\
\ref{f3}), both in the proper FM and AFM cases (all couplings of the same sign)
as well as in those where $J_z$ has the opposite sign of $J_x$. However, when
$|J_z|> |J_y|$ (with (\ref{8}) still holding, e.g.  $J_x< J_y< 0$ and
$J_z>-J_y$), the ellipsoid turns into an hyperboloid and  the N\'eel-type state
{\it ceases to be GS} in the originally AFM case ($J_x$ and $J_y$ negative), as
indicated in the top and bottom right panels. Yet the uniform separable
eigenstate remains  GS  for  {\it both}  $J_x>0$  {\it and} $J_x<0$ (blue lines
in the bottom panels). This is still the case when  $|J_z|$ increases beyond
$|J_x|$, as indicated in the top panels. We just mention that the cases
$J_x>0>J_y>J_z$ and $J_x<0<J_y<J_z$ are equivalent, respectively,  to
$J_z>0>J_y>J_x$ and $J_x>J_y>0>J_z$,  after rotation around the $y$ axis.
Furthermore, cases where  Eq.\ (\ref{8}) does not hold can be transformed to
the present situation by a suitable rotation.

\section{Entanglement in the vicinity of factorization\label{III}}
\subsection{Entanglement in the vicinity of the UGS}
Let us now discuss entanglement in the vicinity of the NTFF $\bm{h}_s$
leading to the uniform GS $|\Theta\rangle$. For simplicity we consider here a
uniform field $\bm{h}$ in a spin $S$ system with TI, where the reduced two-spin
density matrix $\rho_{ij}={\rm Tr}_{\overline{i,j}}|GS\rangle\langle GS|$
(${\rm Tr}_{\overline{i,j}}$ denotes the trace over the complementary
subsystem) depends just on the separation between the two spins. This reduced
state will be in general a mixed state when $|GS\rangle$ is entangled. And such
mixed state is said to be entangled if it cannot be written as a convex mixture
of product states $\rho_i\otimes\rho_j$ \cite{RF.89}, i.e., if it cannot be
generated by local operations and classical communication \cite{NC.00}.

We first show that pairwise entanglement {\it reaches
full range} in the vicinity of the factorizing field $\bm{h}_s$.  \\
{\it Proof}: For $\bm{h}$ close to $\bm{h}_s$, the GS can be obtained by
considering first order perturbative corrections to $|\Theta\rangle$:
\begin{eqnarray}
|GS\rangle&\approx& |\Theta\rangle+{\sum_\nu
\frac{\langle \nu|(\bm{h}-\bm{h}_s)\cdot(\sum_i\bm{S}_i)|\Theta\rangle}
{E_\nu-E_{\Theta}}|\nu\rangle}\nonumber\\
&=&|\Theta\rangle+{\textstyle(\alpha\sum\limits_i{S}_i^{-'}+\sum\limits_{i,j}
\beta_{ij} {S}^{-'}_i{S}^{-'}_{j}+\ldots)|\Theta\rangle}\,,\label{pert}
\end{eqnarray}
where $|\nu\rangle$ are the exact excited eigenstates at $\bm{h}_s$
($H|\nu\rangle=E_\nu|\nu\rangle$, $\langle \nu|\Theta\rangle=0$),  normally
entangled, and ${S}_i^{-'}=R_i S_i^-R_i^\dagger$ the rotated lowering
operators, with $\alpha$, $\beta_{ij}$ and all remaining terms of order $\delta
h_{\perp}$ if $\bm{h}-\bm{h}_s=\delta h_{\perp}\bm{n}_{\theta}^\perp+\delta
h_{\parallel}\bm{n}_\theta$. In the
rotated standard basis $\{\otimes_i|{k'}_i\rangle\}$
(${S}_{z'}^i|{k'}_i\rangle=(S_i-k)|{k'}_i\rangle$) and considering
first $S=1/2$, Eq.\ (\ref{pert}) leads to
\begin{equation}
\rho_{ij}\approx
\begin{pmatrix}1&\alpha & \alpha & \beta_{ij} \\\alpha & 0 & 0 & 0
\\\alpha & 0 & 0 &0  \\ \beta_{ij} & 0 & 0 & 0 \\ \end{pmatrix}+O(\delta h_{\perp}^2)\,.
 \label{red}
\end{equation}
According to the positive partial transposition criterion \cite{P.96,HHH.96}, this
state will be entangled if its partial transpose $\rho^{T_j}_{ij}$ is
non-positive, i.e., if it has at least one negative eigenvalue. But the the
partial transpose of (\ref{red}) has eigenvalues $1,0$ and $\pm\beta_{ij}$ up
to $O(\delta h_{\perp})$, so that $\rho_{ij}$ will be entangled if
$\beta_{ij}\neq 0$. And the exact coefficients $\beta_{ij}$ obtained from
(\ref{pert}) are not strictly zero for any pair $i,j$ linked by successive
applications of the couplings in $H$, due to the two spin excitations present
in the exact eigenstates $|\nu\rangle$.

For higher spins $S$, $\rho_{ij}$ will be more complex (of $(2S+1)^2\times
(2S+1)^2$) {\it but  will still contain a first submatrix of the form
(\ref{red})}. Hence, it will also be entangled if $\beta_{ij}\neq 0$, since the
partial transpose of this block is the first block of the full partial
transpose $\rho_{ij}^{T_j}$, and is non-positive at $O(\delta h_{\perp})$. This
prevents the full $\rho_{ij}^{T_j}$ from being  positive semidefinite (in which
case all principal submatrices should also be so). \qed

For  $S=1/2$, the entanglement between spins $i$ and $j$ can be measured
through the concurrence \cite{W.98} $C_{ij}=2\lambda_{\rm max}-{\rm
Tr}\,M_{ij}$, where $\lambda_{\rm max}$ is the largest eigenvalue of the matrix
$M_{ij}=[\rho_{ij}^{1/2}\tilde{\rho}_{ij}\rho_{ij}^{1/2}]^{1/2}$, with
$\tilde{\rho}_{ij}=\sigma_y\otimes\sigma_y\rho^*_{ij}\sigma_y\otimes\sigma_y$
in the standard basis.  Up to $O(\delta h_{\perp})$, Eq.\
(\ref{red}) then leads  to
\begin{equation}
C_{ij}\approx 2|\beta_{ij}|\propto |\delta h_{\perp}|\,. \label{Conc}
\end{equation}
Note that at this order, $\alpha$ in (\ref{red}) has no effect on the
eigenvalues of $\rho_{ij}^{T_j}$ nor on $C_{ij}$. Eq.\ (\ref{Conc}) implies
that $C_{ij}$, while acquiring finite positive values in the neighborhood of
$\bm{h}_s$, will vanish linearly (as $|\delta h_{\perp}|$) as
$\bm{h}\rightarrow\bm{h}_s$, i.e., as it crosses the straight line of
factorizing fields at a fixed direction $\bm{n}_\gamma$. The corresponding
entanglement of formation \cite{W.98}, $E_{ij}=-\sum_{\nu=\pm}p_\nu\log_2
p_\nu$, with $p_\pm=\frac{1\pm\sqrt{1-C_{ij}^2}}{2}$, is just a convex
increasing function of $C_{ij}$, which vanishes as
$-\frac{1}{4}C^2_{ij}\log_2(C_{ij}^2/4e)$ for $C_{ij}\rightarrow 0$. Hence, for
$\bm{h}\rightarrow\bm{h}_s$ it will vanish essentially as $-\delta
h_{\perp}^2\log_2 |\delta h_{\perp}|$.

It is also seen from (\ref{red}) that the eigenvalues of $\rho_{ij}$ will be
either  $1$ (with negative $O(\delta h^2_{\perp})$ corrections) or small
($O(\delta h^2_{\perp})$). Hence, the entropy $S(\rho_{ij})=-{\rm
Tr}\,\rho_{ij}\log_2\rho_{ij}$, which measures the entanglement between the
pair and the rest of the system \cite{BB.96}, will also vanish essentially as $-\delta
h_{\perp}^2 \log_2 |\delta h_{\perp}|$ for $\bm{h}\rightarrow\bm{h}_s$. The
same behavior at $\bm{h}_s$ will be exhibited by the single spin entropy
$S[\rho(1)]$, where $\rho(1)=\rho_i={\rm Tr}_j\rho_{ij}$ denotes the single
spin reduced state, and also by the  block entropy \cite{VK.03}  $S[\rho(m)]$
of  $m$ contiguous spins, where $\rho(m)$ denotes their  reduced state.
Factorization can in fact be directly seen through the entanglement spectrum
\cite{LH.08,GM.13}, i.e. the set of eigenvalues of the reduced states
$\rho(m)$.  At $\bm{h}=\bm{h}_s$, $\rho(m)$ will have just one nonzero
eigenvalue $p_1=1$, whereas in its vicinity the remaining eigenvalues will
be small, of order $O(\delta h_{\perp}^2)$.

\subsection{Entanglement in the vicinity of the NGS}
We first recall that in the transverse case $\bm{h}=h\bm{n}_z$,  the behavior
of $C_{ij}$ close to the TFF $\bm{h}_{zs}$ in the GS of a finite FM-type  chain
\cite{RCM.08,RCM.09} is different from that described above. Since in the transverse
case the $S_z$ parity $P_z$ is conserved, the exact GS of a finite  spin chain has a
definite parity, exhibiting  parity transitions (the last one at the TFF
$\bm{h}_{zs}$) as the transverse field is increased \cite{RCM.08}. This implies
that for $\bm{h}\rightarrow \bm{h}_{zs}^\pm$, it actually approaches the
entangled definite  parity degenerate side-limits
$|\Theta^{\pm}\rangle=\frac{|\Theta\rangle\pm
|-\Theta\rangle}{\sqrt{2(1\pm\langle-\Theta|\Theta\rangle)}}$, with
$|-\Theta\rangle=P_z|\Theta\rangle$.  These states lead to {\it common} finite
side-limits $C_{\pm}$ of the concurrence $C_{ij}$ for any pair $i\neq j$, given
for $S=1/2$ by \cite{RCM.08}
\begin{equation}
C^\pm=|\frac{\sin^2\theta\cos^{n-2}\theta}{1\pm\cos^n\theta}|\,,\label{C}
\end{equation}
where $n$ is the number of spins  and  $\cos\theta$ is determined  by
(\ref{7}), with $\langle -\Theta|\Theta\rangle=\cos^n\theta$. For finite $n$, a
small but finite discontinuity in $C_{ij}$  is then encountered  as the
transverse field  $\bm{h}$ crosses $\bm{h}_{zs}$, reflecting the parity change
of the GS at $\bm{h}_{zs}$. Of course, exactly at $\bm{h}=\bm{h}_{zs}$, the GS
is two-fold degenerate and entanglement depends on the choice of GS, as in
general degenerate systems \cite{CD.12}. Factorization implies that the minimum
entanglement at this point is zero (obtained when choosing
 $|\pm\Theta\rangle$ as GS), even though the side-limits are finite.

\begin{figure}[t]
 \centering{\hspace*{-.5cm}{\scalebox{.55}{\includegraphics{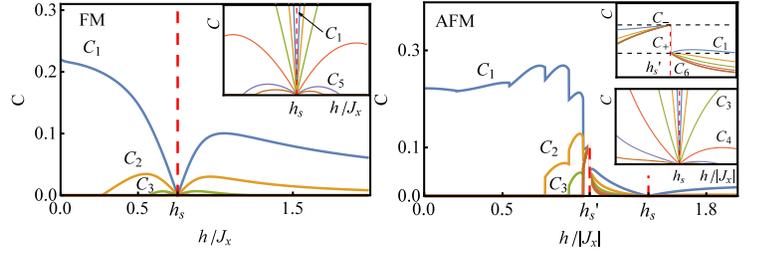}}}}
    \vspace*{-0.5cm}
\caption{(Color online) The concurrences $C_l$ between spins $i$ and $i+l$
(top) in a FM (left) and AFM (right) finite spin $1/2$ $XY$ chain with
$\chi=J_y/J_x=1/2$, as a function of the  magnitude $h=|\bm{h}|$ of the
non-transverse field at fixed orientation
$\bm{n}_\gamma=(\sin\gamma,\cos\gamma)$ in the $xz$ plane, with
$\gamma=0.02\pi$ (FM) and $0.36\pi$ (AFM). For these values there is a single
factorizing field $|\bm{h}_s|\approx 0.76J_x$ (Eq.\ (\ref{10})) in the FM case,
determining a UGS, and two factorizing fields  $|\bm{h}'_s|\approx 1.06 |J_x|$
(Eq.\ (\ref{hgp})) and $|\bm{h}_s|\approx 1.43 |J_x|$ in the AFM case,
corresponding to a NGS and UGS respectively. The insets depict the details in
the vicinity of these fields, showing that all $C_l$'s vanish linearly at
$\bm{h}_s$ (Eq.\ (\ref{Conc})) and approach the finite $l$-independent
side-limits (\ref{Cp}) at $\bm{h}'_s$. All pairs are entangled in the vicinity
of $\bm{h}_s$ and $\bm{h}'_s$, remaining so between both fields in the AFM case
considered. All labels dimensionless. } \label{f4}
\end{figure}

Remarkably, in the AFM chain,  Eq.\ (\ref{C}) {\it remains formally valid  for
the side-limits of $C_{ij}$ at the N\'eel NTFF $\bm{h}'_s$}, i.e., as
$\bm{h}$ at a fixed orientation $\bm{n}_\gamma$ crosses the ellipsoid of
factorizing fields $\bm{h}'_s$. The reason is that the exact GS of a finite
cyclic AFM chain in a uniform field preserves TI away from crossing points and
hence, it approaches for $\bm{h}\rightarrow {\bm{h}'}_s^{\pm}$ the  entangled
TI side-limits
\begin{equation}|\Theta_N^{\pm}\rangle=\frac{|\Theta_N\rangle\pm
 |-\Theta_N\rangle}{\sqrt{2(1\pm \langle-\Theta_N|\Theta_N\rangle)}}\,,
 \label{theN}\end{equation}
where $|\Theta_N\rangle=|\theta_1\phi_1,\theta_2\phi_2,\ldots\rangle$,
$|-\Theta_N\rangle=|\theta_2\phi_2,\theta_1\phi_1,\ldots\rangle=T|\Theta_N\rangle$
denote the degenerate N\'eel-type separable GS's at $\bm{h}'_s$ ($T$ denotes
the one-site translation operator, with $T|\Theta_N^{\pm}\rangle=\pm
|\Theta_N^{\pm}\rangle$). And these states $|\Theta_N^{\pm}\rangle$ lead to
similar  side-limits for the concurrence $C_{ij}$ between any
two spins $i\neq j$ (see (\ref{thm}) below), i.e.,
\begin{equation}
C^\pm=|\frac{\sin^2\theta'\cos^{n-2}\theta'}{1\pm\cos^n\theta'}|\,,\label{Cp}
\end{equation}
where  $\theta'$ is half the difference between the alternating angles of the N\'eel
solution. This angle is determined by \cite{Ku.82}
\begin{equation}
\cos^2\theta'={\textstyle\frac{(J_z+J_y)(J_x+J_y)}{J_x+J_z}\frac{(J_x+J_y)
\cos^2\gamma+(J_z+J_y)\sin^2\gamma}{(J_x+J_y)^2\cos^2\gamma+(J_z+J_y)^2\sin^2\gamma}}\,,
\label{thn}\end{equation}
if  $|\gamma|<\gamma_m$ (Eq.\ (\ref{tang})), as in
the case of Fig.\ \ref{f4} ($\cos^2\theta'$ is given by the  inverse of
(\ref{thn}) if $\gamma_m<\gamma<\pi-\gamma_m$). Hence, in finite chains small
yet finite side-limits together with a discontinuity will be exhibited  by the
concurrences $C_{ij}$ as $\bm{h}$ crosses  $\bm{h}'_s$ at a fixed orientation,
as verified in the top right panel of Fig.\ \ref{f4}.

We can  extend Eq.\ (\ref{Cp}) to general spin $S>1/2$ by still considering the
reduced states $\rho_{ij}^{\pm}$ arising from $|\Theta_N^\pm\rangle$, as those
of two effective qubits, stemming from the single site states
$|\Theta_{Ni}^{\pm}\rangle$, $|\Theta_{Nj}^{\pm}\rangle$, as done in ref.\
\cite{RCM.09} for the TFF. The generalized expression is obtained replacing
$\cos\theta'\rightarrow \cos^{2S}\theta'$ and $\sin^2\theta'\rightarrow
1-\cos^{4S}\theta'$ in (\ref{Cp}). The negativity can be  similarly evaluated
\cite{RCM.09}.

The side-limits at $\bm{h}'_s$ of the reduced state of $m$ given spins,
$\rho^{\pm}(m)$,  can be directly obtained from the exact side-limits
(\ref{theN}) of the full GS. They will be rank 2 mixed states (and not rank one
states, i.e., pure states, as in $\bm{h}_s$), of the form
\begin{eqnarray}&&\rho^\pm(m)=\nonumber\\&&{\textstyle{\frac{|\Theta^m_N\rangle\langle\Theta^m_N|+|-
\Theta^m_N\rangle\langle-\Theta^m_N|\pm (|\Theta^m_N\rangle\langle-
\Theta^m_N|\langle-\Theta^{n-m}_N|\Theta^{n-m}_N\rangle+{h.c.})}{2(1\pm\langle
 -\Theta_N|\Theta_N\rangle)}}\,,}\nonumber\\&&\label{thm}\end{eqnarray}
where $|\pm\Theta^m_N\rangle$ denote the reduced states of the $m$ spins in the
N\'eel states $|\pm \Theta_N\rangle$ and $\langle
-\Theta_N|\Theta_N\rangle=\cos^n\theta'$, $\langle
-\Theta_N^m|\Theta_N^m\rangle=\cos^m\theta'$, with $\cos^2\theta'$ given  by
(\ref{thn}) for $|\gamma|<\gamma_m$. The exact eigenvalues of $\rho^{\pm}(m)$
are $p^{\pm}(m)$ and $1-p^{\pm}(m)$, with
\begin{equation}
p^{\pm}(m)=\frac{(1+\cos^m\theta')(1\pm\cos^{n-m}\theta')}{2(1\pm\cos^n\theta')}
\,. \label{pm}
\end{equation}
The spectrum of $\rho(m)$ will then reduce to these two eigenvalues for
$\bm{h}\rightarrow{\bm{h}'}_s^{\pm}$. These side-limits are {\it independent}
of the choice of the $m$ spins, i.e., the same for $m$ contiguous or separated
spins, as in the UGS of the transverse case \cite{RCM.09}. For general spin
$S$, we should just replace $\cos\theta'\rightarrow \cos^{2S}\theta'$ in
(\ref{pm}).

The ensuing side-limits at $\bm{h}'_s$ of the entanglement entropy $S[\rho(m)]$
are then
\begin{equation}
\!\!\!\!S[\rho^{\pm}(m)]=-p^{\pm}(m)\log_2 p^{\pm}(m)-[1-p^{\pm}(m)]\log_2[1-p^{\pm}(m)]\label{Sm}\,.
\end{equation}
For sufficiently large $m\leq n/2$, the overlap $\langle
-\Theta_N^m|\Theta_N^m\rangle$ vanishes and  $p^{\pm}(m)\rightarrow 1/2$,
$S[\rho^{\pm}(m)]\rightarrow 1$. For $m=2$ we also obtain from (\ref{thm}) the
side-limits of the reduced state of a spin pair, which lead to the separation
independent limits (\ref{Cp}) of the concurrence.

\subsection{Discussion}
\begin{figure}[t]
  \centering{\hspace*{-1cm}{\scalebox{.54}{\includegraphics{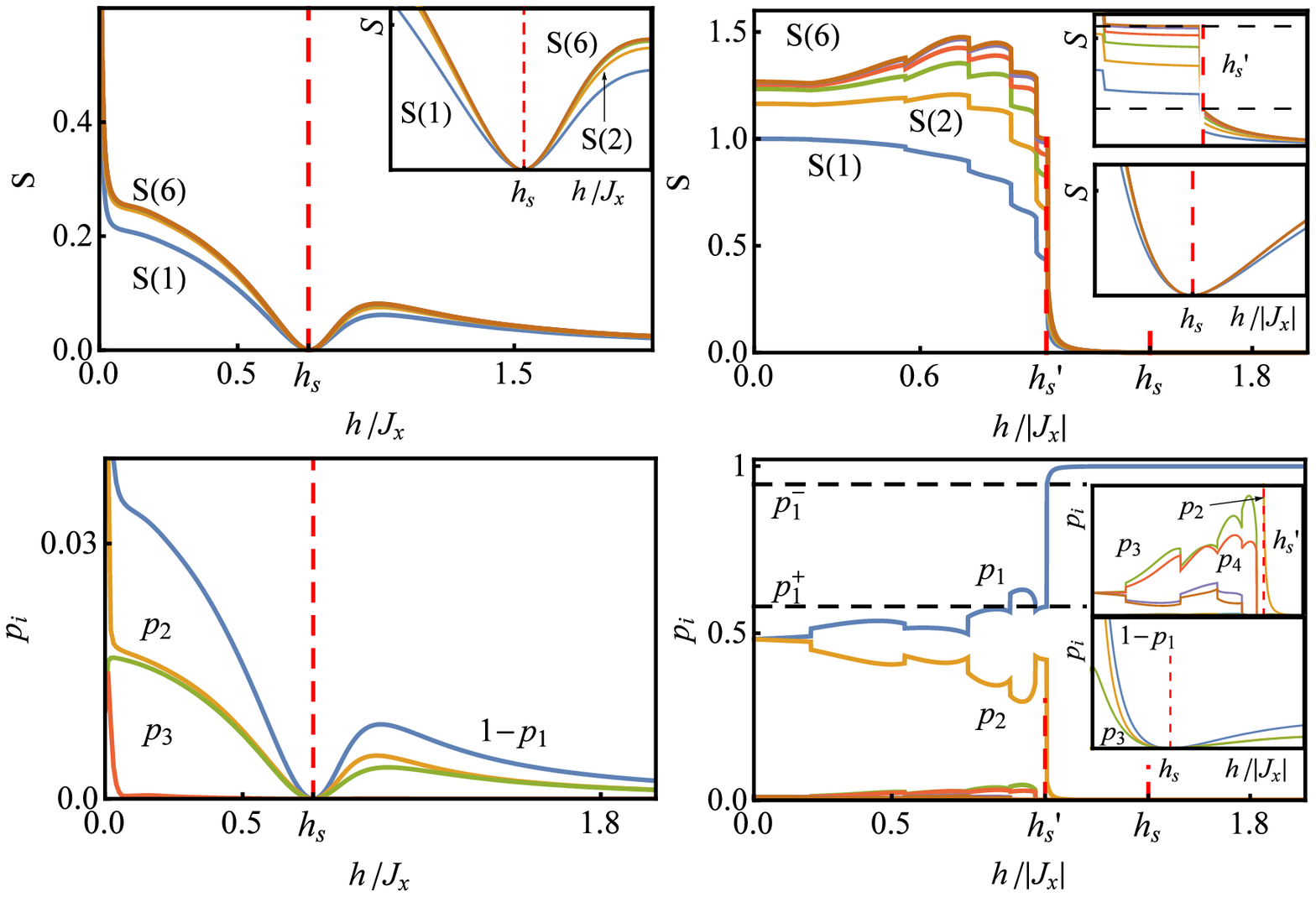}}}}
    \vspace*{-0.5cm}
\caption{The block entropies $S(m)=S[\rho(m)]$ of $m$ contiguous spins (top)
and the eigenvalues $p_i$  (entanglement spectrum) of the corresponding reduced
states $\rho(m)$ for $m=5$, in the same FM (left) and AFM (right) systems of
Fig.\ \ref{f4}. At $\bm{h}_s$ (UGS), $S(m)$ and all but one ($p_1$) of the
eigenvalues of $\rho(m)$ vanish, while for $\bm{h}\rightarrow {\bm{h}'}^\pm_s$
(NGS), $S(m)$ approaches the finite side-limits (\ref{Sm}) (indicated for
$m=n/2$), and two eigenvalues ($p_1$ and $p_2$) remain nonzero, with $p_1$
approaching the indicated side-limits (\ref{pm}). The insets depict again the
details in the vicinity of the factorizing fields.}\label{f5}
\end{figure}

\begin{figure}[t]
  \centering{\hspace*{-0.5cm}{\scalebox{.54}{\includegraphics{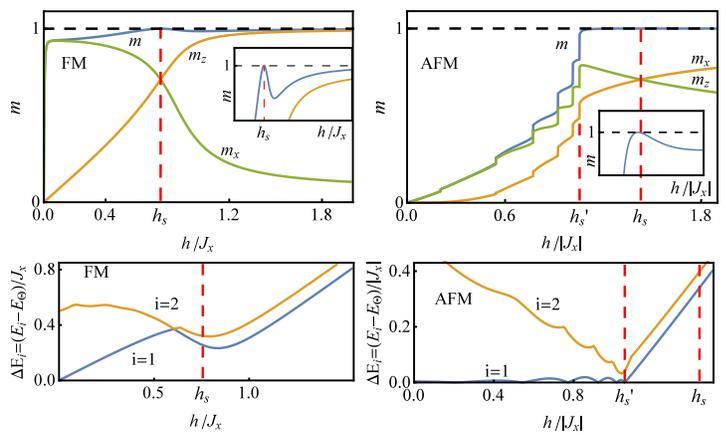}}}}
    \vspace*{-0.5cm}
\caption{(Color online) The scaled intensive magnetizations $m_\mu=\sum_i
2\langle S_i^\mu\rangle/n$ and $m\equiv|\bm{m}|$ (top) and the scaled energy
gap between the ground and the first two excited states (bottom) in the same FM
(left) and AFM (right) systems of Fig.\ \ref{f4}. Notice that $m=1$ at the NTFF
$\bm{h}_s$ determining the UGS, as seen in the insets.  The energy gap shows
that the UGS is well separated from the first excited state, while the NGS is
two-fold degenerate. All labels dimensionless. } \label{f6}
\end{figure}

In Figs.\ \ref{f4}--\ref{f6} we show illustrative exact results for a cyclic FM
(left) and AFM (right) spin $1/2$ chain of $n=12$ spins  interacting through
first neighbor $XY$ couplings ($J_z=0$) with $\chi=J_y/J_x=1/2$, immersed in a
non-transverse field, where all previous effects can be clearly appreciated and
verified. The numerical results were obtained through diagonalization (notice
that an exact analytic solution of the $XY$ chain through the Jordan-Wigner
fermionization \cite{LSM.61} is feasible just for transverse fields \cite{CRM.10}).
All quantities are depicted as a function of the scaled
magnitude $|\bm{h}|/|J_x|$ of the non-transverse field at fixed orientation in
the $xz$ plane ($\gamma=0.02\pi$ in the FM case, $\gamma=0.36\pi$ in the AFM
case). For these orientations there is a single NTFF $\bm{h}_s$ in the FM case,
determining a UGS, whereas in the AFM case there are two NTFF, the first one
$\bm{h}'_s$ corresponding  to a NGS and the second one $\bm{h}_s$  to a UGS.

It is first verified in the top panels of Fig.\ \ref{f4} that while at weak
fields just the first neighbor concurrence $C_1$ is finite in  the present FM
and AFM cases, {\it  all}  concurrences $C_l$ become non-zero in the proximity
of the factorizing fields. As seen in the insets, in the vicinity of $\bm{h}_s$
their behavior is correctly described by Eq.\ (\ref{Conc}), vanishing all
linearly with $|\bm{h}-\bm{h}_s|$  for $\bm{h}\rightarrow \bm{h}_s$
($\beta_{ij} \propto\delta h_{\perp} \eta^{-{|i-j|}}$ ($\eta>1$) in the case of
Fig.\ \ref{f4}, changing sign as $\bm{h}$ crosses $\bm{h}_s$).  On the other
hand, in the AFM case they all approach  the finite $l$-independent distinct
side-limits (\ref{Cp}) at $\bm{h}'_s$ (here $\cos\theta'\approx 0.92$ and
$C^-\approx0.11$, $C^+    \approx 0.049$). Both factorizing fields appear
successively as the field increases along orientations
$\theta<\gamma<\gamma_{m}$, leading to a rather broad interval of
``long-range'' pairwise entanglement located between $\bm{h}'_s$ and
$\bm{h}_s$, as appreciated in the right panel.  It is also seen that  all $C_l$
exhibit jumps for $|\bm{h}|<|\bm{h}'_s|$, the last one at $\bm{h}'_s$, which
reflect the $n/2$  translational parity transitions of the exact GS,  as
discussed below. We remark that while the  side-limits (\ref{Cp}) at
$\bm{h}'_s$ diminish as the number $n$ of spins increases, the  finite  values
of the  $C_l$'s  in the vicinity of both $\bm{h}'_s$ and $\bm{h}_s$  persist
for larger sizes.

The block entanglement entropies of $m$ contiguous spins are depicted in Fig.\
\ref{f5}. Block entropies in $XY$ or $XYZ$ spin  chains have been studied in
detail just for zero or transverse fields \cite{VK.03,IK.05,GM.13}, including
also block Renyi entropies \cite{FIK.08,EE.11,GM.13}. It is first verified that
the von Neumann entropies $S[\rho(m)]$  vanish essentially quadratically in the
vicinity of $\bm{h}_s$, whereas in the AFM case they approach the finite
side-limits (\ref{Sm}) at the N\'eel factorizing field $\bm{h}'_s$ (here
$S[\rho^+(m)]\approx 0.31$ while $S[\rho^-(m)]=1$ for $m=n/2$; notice from
(\ref{pm}) that $p^-(m)=1/2$ for $m=n/2$ $\forall$ $n$). In the FM case these
entropies rapidly saturate as $m$ increases for all non-zero fields, showing
then a non-critical behavior, whereas in the AFM case, while above $|\bm{h}_s|$
they become  small  ($<0.01$)  and also rapidly saturate, below $|\bm{h}'_s|$
they are larger and show an appreciable dependence with block size.

The behavior of these entropies can be better understood by means of the
entanglement spectrum, shown in the bottom panels, where the eigenvalues $p_i$
of $\rho(m)$ for $m=5$ are depicted. Results for other $m>1$ are similar. In
the FM case, there are three dominant eigenvalues, with $p_1$ close to $1$, and
the behavior of $p_2$ and $p_3$ resembles that of $S(m)$:  All eigenvalues
except $p_1$ vanish (quadratically) at $\bm{h}_s$.  However, in the AFM case it
is seen that for $|\bm{h}|<|\bm{h'}_s|$, both $p_1$ and $p_2$ are significant
and comparable, indicating roughly an approximate rank $2$ reduced state. When
$\bm{h}\rightarrow {\bm{h}'}_s^\pm$, $\rho(m)$ becomes exactly a rank $2$ state
and just $p_1$ and $p_2$ are nonzero, in agreement with Eqs.\
(\ref{pm})--(\ref{thm}). The behavior is similar to that observed for a
transverse field \cite{GM.13}. As expected, the GS transitions taking place in
this sector are clearly visible in both the block entropy and the entanglement
spectrum. For $|\bm{h}|>\bm{h'}_s$ there are just three dominant eigenvalues,
with $p_1$ much larger than the rest, as in the FM case. All but $p_1$ vanish
again at the second factorizing field $\bm{h}_s$.

Let us remark that at zero field, the results for any entanglement measure in
the FM and AFM  $XY$ chains with first neighbor couplings are strictly coincident, 
since the corresponding Hamiltonians can be transformed into each other by a local 
rotation of angle $\pi$  around the $z$ axis at all even sites, which does not 
affect entanglement measures. This fact explains the pronounced increase of the 
block entropies of the FM case as the field vanish, since they approach in this 
limit the higher AFM values. This symmetry no longer holds for finite fields not 
pointing along the $z$ axis.

In spin $1/2$ systems, the magnetization can be used as a {\it separability
witness}: The quantity $m=2\sum_i|\langle\bm{S}_i\rangle|/n$ satisfies  $m<1$
in any pure entangled state of such system, with $m=1$ if and only if the pure
state is completely separable. For a state with TI, $m=2|\bm{M}|/n$, with
$\bm{M}=\langle \sum_i \bm{S}_i\rangle$ the total magnetization. Hence $m=1$ at
the NTFF ${\bm h}_s$, as verified in the top panels of Fig.\ \ref{f5},
entailing a non-monotonous behavior of $m$ for increasing fields, as seen in
the insets. We have numerically checked that such non-monotonous behavior
persists for larger sizes, indicating that it is not a finite size effect.
Therefore, through a careful measurement of $\bm{M}$ or the associated
susceptibility as a function of the applied field, one could  be able to
identify the NTFF $\bm{h}_s$.

In the bottom panels of Fig.\ \ref{f5} it is seen that the UGS at $\bm{h}_s$ is
non-degenerate and well separated from the first excited state, whereas the NGS
at $\bm{h}'_s$ is two-fold degenerate. Actually, as seen from  the energy gap
and also from the magnetization and previous entanglement measures,  while no
transitions are observed in the FM case, in the AFM case the exact GS exhibits $n/2$
transitions as $|\bm{h}|$ increases at fixed $\gamma$, the last one taking
place at $\bm{h}'_s$. They correspond to ``translational parity'' transitions
$|GS^{\pm}\rangle\rightarrow |GS^{\mp}\rangle$, with $|GS^{\pm}\rangle$ the
exact TI ground states, which satisfy $T|GS^{\pm}\rangle=\pm|GS^{\pm}\rangle$
($T$ is the one-site translation operator).  These transitions are similar to
those observed for transverse fields in both AFM or FM systems
\cite{RCM.08,RCM.09,CRM.10,GM.13}, where they are related with spin parity
transitions and also end at the corresponding transverse factorizing field
\cite{RCM.08,RCM.09,CRM.10}. Hence, $\bm{h}'_s$ still represents, in the
non-transverse case, a critical field for the finite system, indicating the
passage to a different regime.

\begin{figure}[t]
  \centering{\hspace*{-0.5cm}{\scalebox{.54}{\includegraphics{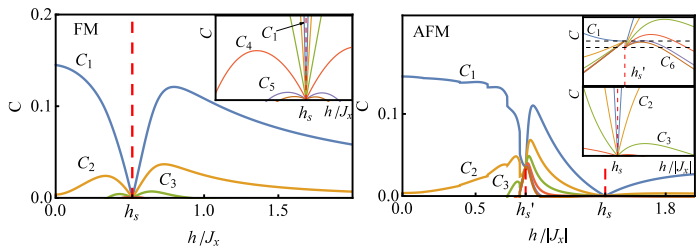}}}}
    \vspace*{-0.5cm}
\caption{(Color online) The concurrences $C_l$ between spins $i$ and $i+l$ in
FM and AFM $XYZ$ chains with $J_y/J_x=1/2$,  $J_z=0.2 |J_x|$ and  $J_x>0$
($<0$) in the left (right) panel. The orientation of the applied magnetic field
in each panel is the same as that of Figs.\ \ref{f4}--\ref{f6}. The factorizing
fields are now $|\bm{h}_s|\approx 0.52 J_x$ in the FM case, where $\chi=0.375$, and
$|\bm{h}_s|\approx 1.39 J_x$, $|\bm{h}'_s|\approx 0.84J_x$ in the AFM case, where
$\chi=0.583$. The side-limits (\ref{Cp}) at $\bm{h}'_s$ and the linear vanishing
of all concurrences at $\bm{h}_s$ are again verified.} \label{f7}
\end{figure}

Finally, we depict in Fig.\ \ref{f7} results for the pairwise concurrence in a
chain with full $XYZ$ couplings, a system which cannot be mapped to independent
fermions even in the transverse case (\cite{LSM.61,EE.11}). We have set
$J_z=0.2|J_x|$, in both the FM ($J_x>J_y>0$) and AFM ($J_x<J_y<0$) cases, using
the same previous field orientations. The behavior is quite similar to that of
Fig.\ \ref{f4}, with the GS translational parity transitions also present in
the AFM case. One  just notes the higher values of $C_l$ above the factorizing
fields in both cases, and the closer side-limits at $\bm{h}'_s$ in the AFM case
(now $C^-\approx 0.036$, $C^+\approx 0.032$), due to the different value of the
anisotropy ratio $\chi$. The values at zero field are again still strictly
coincident due to the same sign of $J_z$. Results for the block entropy and
entanglement spectrum for the finite case considered are also qualitatively
similar to the previous results.

\section{Conclusions}
We have first determined the general conditions for the existence of separable
eigenstates with maximum spin at each site in general arrays with anisotropic
$XYZ$ couplings immersed in a  non-transverse field. The set of factorizing
fields can be characterized by the local fields orthogonal to the local
alignment direction, plus arbitrary fields parallel to the latter. We have next
identified the possibility of a uniform non-degenerate  separable GS in quite
general systems of arbitrary spin, including FM and AFM-type chains and arrays,
for fields parallel to a principal plane (Fig.\ \ref{f2}). The coupling range
can be arbitrary, provided the anisotropy ratio $\chi$ is constant. In AFM
$XYZ$ chains with first neighbor couplings, this separable solution coexists in
field space with the N\'eel-type separable solution.

We have also demonstrated, for arbitrary spin, that pairwise entanglement
reaches full range in a finite array in the vicinity of the factorizing field
determining the uniform solution, with the concurrence vanishing linearly at
this field. Full range is also reached at the N\'eel NTFF, although here it was
shown that in finite cyclic even chains, the pairwise concurrence reaches
finite side-limits in its vicinity, which were analytically evaluated. This
NTFF was shown to correspond to the last parity transition of the GS in the
finite cyclic chain.  Block entropies were also analyzed and shown to vanish
quadratically at the uniform NTFF, while reaching again finite (and
analytically determined) side-limits at the N\'eel NTFF in these finite chains.

Present results and limits are also applicable to more complex systems, like
dimerized chains and arrays \cite{G.09,CRM.10,BRCM.15}. The recent possibility
of performing quantum simulations of spin chains and lattices with tunable
couplings  through cold atoms in optical lattices \cite{S.11,LSA.12,GAN.14} or
trapped ions \cite{GAN.14,B.12,K.09,BR.12,SK.12} augments the potential of the
present results. Such experiments could then provide valuable insights into the
remarkable phenomenon of factorization and its relation with entanglement and
criticality in finite many body systems.

\begin{acknowledgments}
The authors acknowledge support of CONICET (MC, NC) and CIC (RR) of Argentina.
\end{acknowledgments}

\end{document}